\documentclass[graybox,natbib]{svmult}
\bibpunct{(}{)}{;}{a}{}{,}
\usepackage{helvet}
\usepackage{courier}
\usepackage{makeidx}
\usepackage{graphicx}
\usepackage{multicol}
\usepackage[bottom]{footmisc}
\usepackage[normalem]{ulem}
\usepackage{hyperref}
\usepackage{soul}
\usepackage[margin=1in]{geometry}

\usepackage{lmodern}

\newcommand{\hbindex}[1]{#1}
\newcommand{\todo}[1]{ }
\makeindex

\newcommand{\ergscm}{erg s$^{-1}$ cm$^{-2}$}

\begin{document}

\title*{The Hot Jupiter Radius Anomaly and Stellar Connections}
\author{Daniel P. Thorngren}
\institute{Daniel P. Thorngren \at Johns Hopkins University, 3400 N. Charles Street, Baltimore, MD 21218, USA, \email{dpthorngren@jhu.edu}}
\maketitle
\vspace{-3cm}

\abstract{The extremely close proximity of hot Jupiters to their parent stars has dramatically affected both their atmospheres and interiors, inflating them to up to twice the radius of Jupiter.  The physical mechanism responsible for this inflation remains unknown, though many proposals have been put forward.  This chapter reviews the known hot Jupiter population, the proposed inflation mechanisms, and the evidence for and against them collected thus far.  In doing so, it covers the ways that hot Jupiter interiors may be simulated computationally in detail and present some useful formulas for estimating their radii, heating, intrinsic temperature, and tentative magnetic field strength.  It also covers the related issues of hot Jupiter intrinsic temperatures and radiative-convective boundaries, the potential connection with planetary magnetic fields, and the effects of stellar tides on the planet.  Finally, it concludes with the suggestion that more than one mechanism may be operating in concert with each other and propose various avenues for future progress in understanding these objects.}

\section{Introduction}
The discovery of \hbindex{gas giant planets} on close-in orbits around their parent stars \citep[the first being 51 Pegasi b;][]{Mayor1995} came as something of a surprise and raised the question of how such objects could be formed \citep[e.g.,][]{Pollack1996, Ida2004}.  Some argued the planets could have formed in-situ \citep[e.g.,]{Bodenheimer2000}, while others favored migration from wider separations due to interactions with the disk \citep[e.g.,][]{Masset2003} or the star \citep[eccentric migration][]{Marzari2002, Ford2006}.  This was the first of two great problems in \hbindex{hot Jupiter} physics.  The second emerged when the transit method revealed a radius for HD 209458 b: $1.27 \pm 0.02~R_J$ \citep{Charbonneau2000} and $1.42\pm0.1~R_J$ \citep{Henry2000}.  This large size stood in stark contrast with our previous understanding of giant planet physics, which predicted a radius of $\sim1.1~R_J$ \citep{Bodenheimer2001,Guillot2002}.

Initially, this discrepancy was understood to be either the result of observational and theoretical uncertainties \citep{Burrows2003, Burrows2004} or as an exceptional case of a large planet \citep{Bodenheimer2003} having an unusually hot interior due to tidal interactions with the parent star \citep{Bodenheimer2001}.  \cite{Guillot2002} proposed that while the planet indeed must be very hot \citep[which was supported by][]{Baraffe2003,Chabrier2004}, the most likely source of heating was not tidal interactions but rather the intense instellation from the parent star heating the interior through a previously unknown physical mechanism.  A small number of additional transiting hot Jupiters were discovered, including OGLE-TR-56 b \citep{Konacki2003} and TrES-1 \citep{Alonso2004}, revealing more moderately inflated hot Jupiters similar to HD 209458 b.

\begin{figure}[t]
    \centering
    \includegraphics[width=\textwidth]{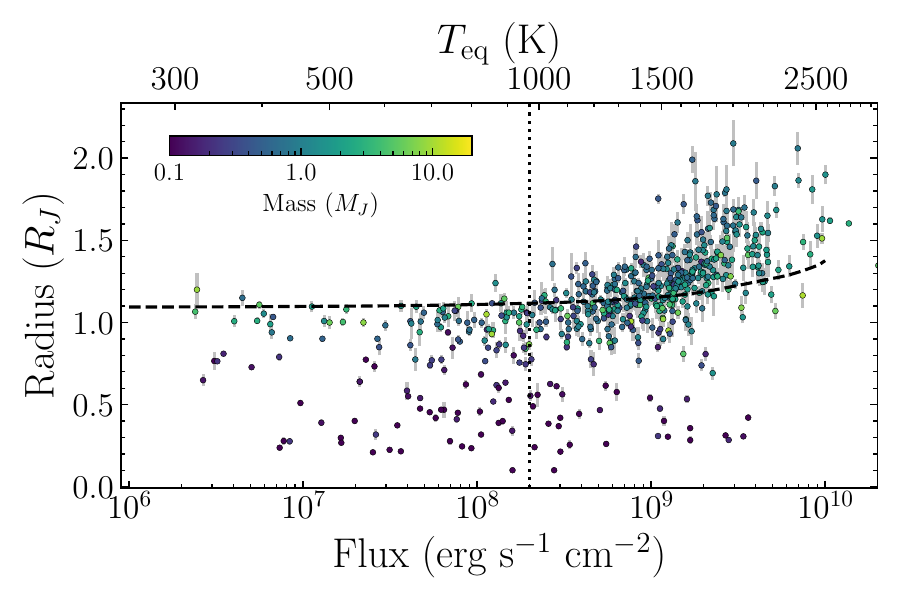}
    \caption{The radius of hot (right of the vertical line at $2\times10^8$ erg s$^{-1}$ cm$^{-2}$) and warm Jupiters (left of the line) as a function of incident flux, or equivalently, equilibrium temperature for zero albedo and full heat redistribution.  Planets are colored by their mass.  The horizontal dashed line is the radius of a 4.5 Gyr old, 1 $M_J$, pure H/He planet (no metals), with no added heating, which is a rough upper-limit on evolved, unheated planets.  Note that hot Jupiters are systematically larger than this limit, that their excess radius is correlated with flux, and that this effect is lessened for higher-mass planets.}
    \label{fig:fluxRadius}
\end{figure}

The number of known hot Jupiters grew substantially in the late 2000s thanks in large part to discoveries made by the Hungarian Automated Telescope \cite[HAT;][]{Bakos2004, Bakos2007} and Wide Angle Search for Planets \citep[WASP;][]{Pollacco2006, CollierCameron2007} surveys.  These exhibited consistently larger radii than theoretical models could produce (Fig. \ref{fig:fluxRadius}), demonstrating that the radius anomaly was not a special case for the first few hot Jupiters, but instead a process broadly affecting the entire population.  Subsequent discoveries using Kepler \citep{Borucki2010}, K2 \citep{Howell2014}, CoRoT \citep{Auvergne2009}, HAT-South \citep[HATS; ][]{Bakos2013}, the Transiting Exoplanet Survey Satellite \citep[TESS;][]{Ricker2014}, and others have expanded the number of transiting hot Jupiters with mass measurements to over 500!  This chapter will give an overview of this population, explain how interior structure models of these objects are calculated, list some of the proposed explanations for their radii, and identify important observational implications of these hypotheses.

\section{The Observed Population}
The current set of giant exoplanets with known radii and masses is shown in Fig. \ref{fig:fluxRadius}, plotted as radius against flux.  The black dashed line depicts the radius of a pure H/He object with no extra internal heating.  As originally observed by \citet{Miller2011} and \citet{Demory2011}, radius inflation can be seen to turn on at $2\times10^8$ erg s$^{-1}$ cm$^{-2}$ ($\sim1000$ K), marked with a vertical dotted line.  Above this radii are correlated with the incident flux.  The most highly inflated planets reach $2~R_J$ at \hbindex{equilibrium temperatures} in the 1500-1800 K range, and are generally somewhat less massive than Jupiter.  On average, more massive planets are smaller for a given incident flux.

The observed hot Jupiters population predominantly orbits F, G, and K dwarf stars due to a combination of observation bias against their detection around high-mass stars (the transit depth is much lower) and an inherently low occurrence rate around M dwarfs \citep{Johnson2010}.  This variation in stellar mass and therefore luminosity allows us to examine whether the radii are most strongly correlated with the incident flux (most related to atmospheric effects) or the orbital period (more related to tidal effects).  \citet{Weiss2013} finds that the incident flux is a \emph{much} better predictor of the radius anomaly than the period.  They also provide another useful result: an empirical mass-flux-radius relation,
\begin{equation}
    \left(\frac{R}{R_\oplus}\right)= 2.45 \left(\frac{M}{M_\oplus}\right)^{-0.039} \left(\frac{F}{\mathrm{erg~s^{-1}~cm^{-2}}}\right)^{0.094}
\end{equation}
for planets with $M>150~M_\odot$.  $R$ is the planetary radius in $R_\oplus$, $M$ is the planetary mass, and $F$ is the incident flux onto the planet in \ergscm.

For completeness and comparison, a few other empirical fits to hot Jupiter radii bear mentioning.  \citet{Chen2017} provide a rigorously-derived mass-radius relation for a wide range of planetary masses and an easy-to-use code for calculating predictions in either direction, but note that they do not use the incident flux as a predictor. They therefore output radii corresponding with a typical flux of around $10^9$ \ergscm.  For hot Jupiters in particular, \citet{Thorngren2021} experimented with a large number of functional forms with a larger sample size but recommends a similar form to that of \citet{Weiss2013}:
\begin{equation}\label{eq:massRadiusFluxThorngren}
    R = 1.22 M^{-0.042} (F/10^9)^{0.137 - 0.072 \log_{10}(M)}
\end{equation}
where $R$ is the radius in $R_J$, $M$ is the mass in $M_J$, and $F$ is the incident flux in \ergscm.  Note the addition of a cross-term between flux and radius: the inflationary effect of flux decreases with increasing mass.  

In some cases it may be desirable to estimate the radius in the absence of anomalous heating -- e.g. to calculate the radius anomaly as in \citet{Laughlin2011}.  \citet{Fortney2007} calculated a grid of model radii that may be interpolated for this purpose.  Alternatively, \citep{Thorngren2019a} estimate the following \hbindex{mass-radius relation} for giant planets with $F < 2\times10^8~\mathrm{erg~s^{-1}~cm^{-2}}$:
\begin{equation}
    R_\mathrm{uninflated} = 0.96 + 0.21\log_{10}(M)-0.2\log_{10}(M)^2 \pm 0.12 M^{0.215}
\end{equation}
where $R_\mathrm{uninflated}$ is the planetary radius in $R_J$ and $M$ is the planetary mass in $M_J$.  The subscript ``uninflated" reminds readers that this is the planet radius \emph{assuming no inflation}, based on the warm giant planets.  Note that a predictive error is also estimated as a function of mass.

Finally, it is worth noting a few hot Jupiters of particular interest.  HD 209458 b was the first discovered hot Jupiter, but it also is a fairly typical example of the class that is extremely well-studied, making it a good choice if readers want a single "typical case" to study.  KELT-9 b \citep{Gaudi2017} is the hottest known hot Jupiter with an impressive $T_\mathrm{eq} = 4600$ K, rivaling some K stars in surface temperature and thereby hosting an atmosphere with radically different chemistry from typical hot Jupiters \citep[e.g.,][]{Hoeijmakers2018}.  The largest planet known is one of WASP-17 b, WASP-78 b, or WASP-79 b, all hot Jupiters of about 2 $R_J$ whose radius uncertainties prevent us from knowing which is largest.  WASP-12 b, WASP-121 b, and WASP-103 b are believed to be the most tidally-distorted planets \citep{Wahl2021}, though by their radius vs Roche radius WASP-19 b and Qatar-10 b may be comparably distorted.  Lastly, the shortest-period hot Jupiter with well-determined properties is currently TOI-2109 b \citep{Wong2021}, orbiting its star in just 16 hours.

\subsection{Observational Biases} \label{sec:samplingBias}
This section has examined the full sample of known transiting hot Jupiters, but one should bear in mind the observational biases that might be affecting it.  Primarily, this refers to the biases of the transit method, whereby larger objects produce more easily detectable transit signals.  Hot Jupiters are sufficiently large and short-period that for bright stars they are extremely likely to be successfully observed, limiting the importance of this bias.  Indeed, \citet{Yee2021} found that for stars with $G < 10.5$, the \hbindex{completeness} (excluding the poles and galactic plane) is statistically nearly 100\%!  Further, they argue that TESS might reasonably expand this completeness to a limiting G-magnitude of about 12.5 \citep[see also][]{Sullivan2015, Zhou2021}.  This will present an exciting opportunity to better understand these planets, but one should still remember that the order in which they are discovered will not be random; the selection function for follow-up is undoubtedly complex, and unusual planets may be prioritized for confirmation.  For planets that are observed, the large pixels of TESS \citep{Ricker2014} could risk light curve contamination from background or companion stars (causing an underestimation of the planetary radius), but these cases can be caught and corrected for through either spectroscopy of the star or by comparison with the GAIA data \citep[e.g.][]{Mugrauer2020}.

Another characteristic that biases the observed population of hot Jupiters is the subtle dependence on the age of the parent star.  The principal cause of this is \hbindex{tidal inspiral} \citep{Hamer2019}, in which the closest-in hot Jupiters raise tides in their host stars that saps them of their angular momentum, decaying their orbits and eventually causing \hbindex{Roche lobe overflow} \citep{Valsecchi2014,Valsecchi2015}.  The tidal circularization of eccentric planets into hot Jupiters is expected to be generally very fast \citep{Jackson2008}, with a few rare cases of planets being observed presently undergoing this process, such as HD 80606 b \citep{Naef2001}.  Finally, while higher-mass giant planets are thought to be fairly resistant to \hbindex{XUV-driven mass loss} \citep{Murray-Clay2009, Dwivedi2019}, planets near Saturn's mass or lower can still undergo substantial XUV-driven mass loss \citep{Caldiroli2022,Kurokawa2014,Thorngren2023}.  As such, hot Jupiters orbiting young stars \citep[e.g.,][]{Newton2019,Zhou2021} are a valuable point of comparison from before these processes have completed.

\begin{figure}[t]
    \centering
    \includegraphics[width=\textwidth]{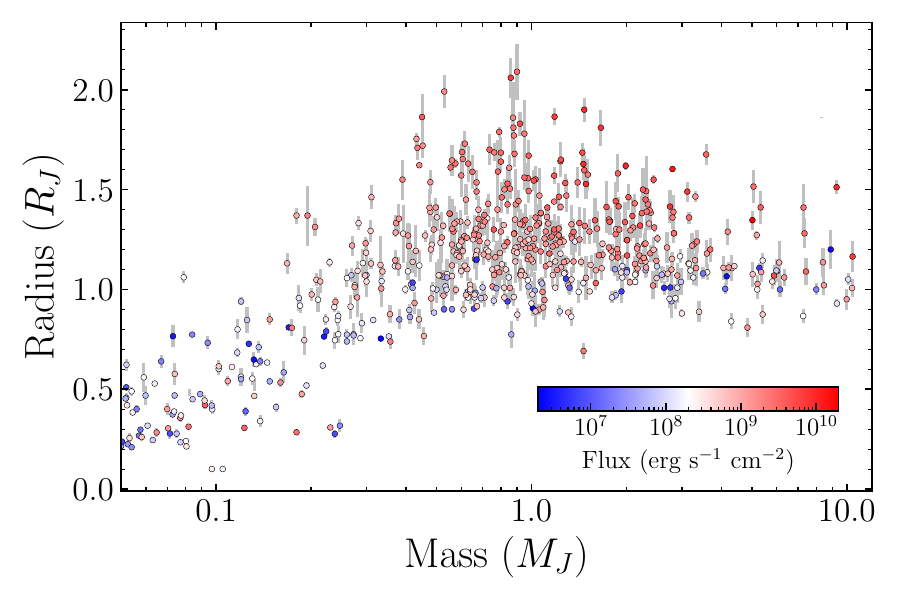}
    \caption{The radius of the giant planet population vs their mass, colored by the incident flux such that blue points indicate planets below the hot Jupiter inflation threshold and red points are for those above it.  At around 0.3 $M_J$, electron degeneracy pressure dominates the mass-radius relation of the cooler giants, leveling off their radius with mass.  For hot Jupiters the flux strongly influences the radius, most strongly for planets in the 0.5 to 2 $M_J$ range.  The radii of planets below about 0.2 $M_J$ do not show correlation with either flux or mass.}
    \label{fig:massRadius}
\end{figure}

\section{Modeling Hot Jupiter Interiors}
The radius of a giant planet is mainly determined by the thermal state and composition of its interior, with only a small portion ($\sim1\%$, depending on the temperature) of the radius contributed by the atmosphere.  As such, modeling these interiors is the first step in understanding the radius inflation problem.  Even very modern computers are not fast enough to run fluid-dynamical simulations in three dimensions over billion-year timescales, so the long-term evolution of giant planets is relegated to one-dimensional models.  This section will review how one may construct such models of giant planets at a single point in time (static models), how these may be evolved forward through time, and additional relevant physics that one may wish to include in modeling these objects, depending on circumstance.

It should be mentioned that there are existing software packages that can be used to model \hbindex{planetary interiors}, the most prominent being MESA \citep{Paxton2013}.  While originally built for modeling the interiors and evolution of stars, the fundamental PDEs (partial differential equations) being solved are the same, allowing it to be successfully adapted to gas giant use cases on numerous occasions \citep[e.g.,]{Chen2016,Mankovich2016}.  Still, prospective users should be aware that these modifications are non-trivial.  Another open-source option is MAGRATHEA \citep{Huang2022}; while mainly focused on the simulation of terrestrial or sub-Neptune planets, it can also be used for the simulation of true gas giant interiors.

\subsection{Static Models}
At an instant in time, the state of a giant planet can be described by its total mass $M$ and a complete description of the interior composition and temperature at a given mass shell $m$.  For a one-layer fully-convective planet, the composition can be the metal mass-fraction of the planet $Z_p$: 0 representing a metal-free planet and 1 being a planet composed entirely of metal (no hydrogen or helium).  Due to the assumption of a fully-convective planet, the thermal state can be described by the specific entropy $s$.  An equation of state maps the specific entropy to a temperature for any given pressure, and the curve through temperature-pressure space is known as the adiabat.  For giant planets, the \hbindex{specific entropy} of a hydrogen-helium envelope ranges from 5 (very ``cold") to 12 (very ``hot") $k_b$ per baryon, where $k_b$ is the Boltzmann constant.  Giant planets begin their life with specific entropy somewhere in the 8-12 $k_b$ per baryon range \citep{Marley2007,Cumming2018}, which then decreases as energy is released through the atmosphere.

Given the mass, composition, and temperature structure of the planet, its radius can be calculated by solving the equations of hydrostatic equilibrium, mass conservation, and an appropriate \hbindex{equation of state} (EOS):

\begin{eqnarray}
    \frac{dP}{dm} = -\frac{G m}{4 \pi r^4}\\
    \frac{dr}{dm} = \frac{1}{4 \pi r^2 \rho}\\
    \rho^{-1} = \sum_i f_i / \rho_i(s,P) \label{eq:additiveVolumes}
\end{eqnarray}

Eq. \ref{eq:additiveVolumes} shows the \hbindex{additive volumes approximation} across the equations of state of multiple components with mass fraction $f_i$ and density $\rho_i$ as a function of pressure and specific entropy.  For these, one must decide what materials are to be modeled.  A typical choice would start with hydrogen and helium at the solar \citep[see][]{Asplund2009} or protosolar ratios.  For the latter, \citet{Serenelli2010} give $Y=.278\pm0.006$ and $Y=.273\pm0.006$, depending on the exact assumptions made.  With a middling $Y=.275$ and $Z=.0142$ \citep{Asplund2009}, we have $Y/X=0.387$.  To obtain the density of H/He one should interpolate from an EOS table containing densities at various pressures and specific entropies.  \citet{Chabrier1992} was the dominant choice for over a decade, but more recent ab-initio calculations \citep{Nettelmann2008, Militzer2013, Chabrier2019} have supplanted it.  \citet{Chabrier2021} would be a strong choice for H/He as of the time of writing, as it combines information from both \cite{Militzer2013} and \citet{Chabrier2019}.

\begin{figure}[t]
    \centering
    \includegraphics[width=\textwidth]{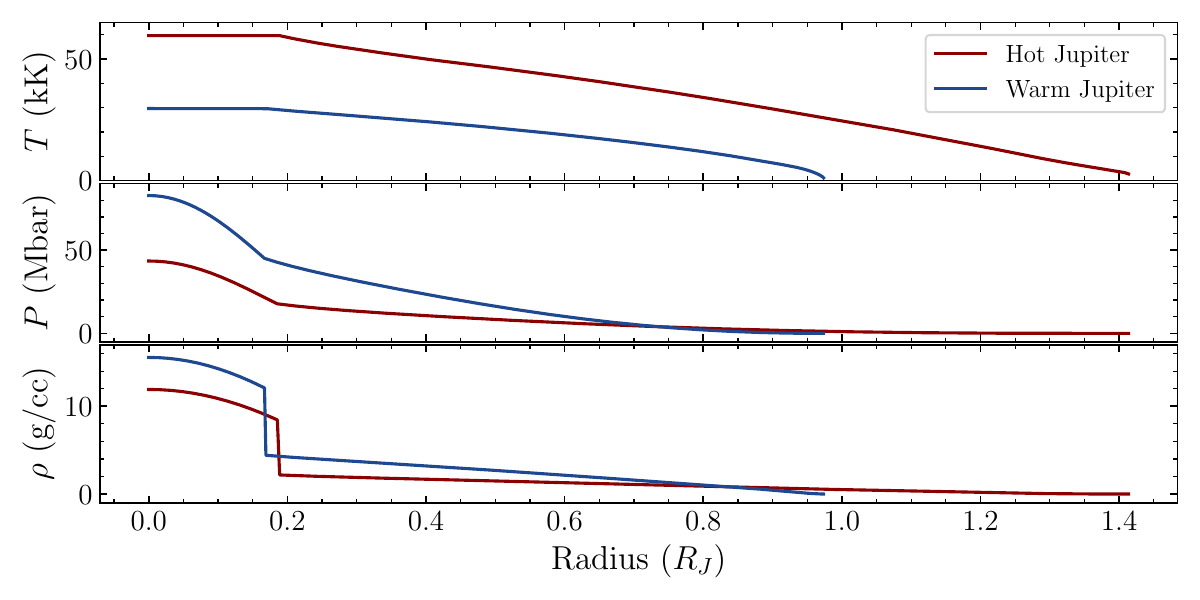}
    \caption{The temperature, pressure, and density of a 1 $M_J$ planet with a 15 $M_\oplus$ core and an envelope metallicity $Z_p=0.1$ at specific entropies of 9 and 7 $k_b$ per baryon.  These roughly correspond to a hot Jupiter at $T_\mathrm{eq}\sim1500$ K and an evolved warm Jupiter.  The cooler interior of the warm Jupiter has a higher density at a given pressure, which in turn also increases the pressures, compressing even the non-thermally-expansive core.  The result is much smaller planet: $0.97~R_J$ vs $1.41~R_J$.}
    \label{fig:staticExample}
\end{figure}

Gas giants interiors also contain metals, which are typically represented in the equation of state by water and/or a representative rock (often olivine, since it is a large portion of the Earth's mantle).  The outer solar system icy moons have roughly 50-50 ice-to-rock compositions, so this ratio is a reasonable assumption for the metal of gas giants formed outside an ice line.  That said, at the extreme pressures found in gas giants, the difference in density between these various materials isn't very large, so this choice is a second-order effect on the observables (radius, luminosity).  The offerings for metal EOSs are more sparse than for H/He, but options include ANEOS \citep{Thompson1990}, SESAME \citep{Lyon1992}, H2O-REOS \citep{Nettelmann2010}, \citet{Mazevet2019a}, and AQUA \citep{Haldemann2020}.  MAGRATHEA \citep{Huang2022} also provides a fair number of analytic equations of state for metals and a table-output function.

\runinhead{Multilayer Planets}
The discussion of the previous section has implicitly assumed that the planet in question is of uniform composition and convective throughout, but this is not necessarily the case.  A major issue with this is that the core accretion model of planet formation requires a heavy-element core to form first in order to reach runaway gas accretion \citep{Pollack1996}.  This has led to the use of a two layer model consisting of an isothermal metallic core and a convective H/He envelope \citep[e.g.,][]{Marley2007, Lopez2014, Piso2015}, especially in the sub-Neptune regime and for studies focused on the gas-accretion process \citep[e.g.,][]{Lee2014}.  Fig. \ref{fig:staticExample} shows an example interior model for a hot and warm Jupiter featuring an envelope surrounding a distinct rock-ice core.

For Uranus and Neptune, these require at least the presence of an additional water-rich envelope layer between the core and outer envelope \citep[see][]{Fortney2010a, Bailey2021}.  Jupiter and Saturn are both believed to feature ``fuzzy" cores \citep{Wahl2017, Debras2019, Movshovitz2020, Mankovich2021}, which is to say that the composition gradually transitions from H/He poor to rich, rather than via a sharp boundary.  This gradual transition may feature a staircase of \hbindex{layered semiconvection}, in which convective layers are separated by thin non-convecting boundaries with steep temperature and composition gradients \citep{Leconte2012}.  Such a staircase appears to be the natural outcome of initial composition gradients \citep{Moll2016, Helled2017, Fuentes2022}, and could have a significant impact on the thermal evolution if the number of steps is large enough \citep{Chabrier2007,Leconte2012}.  On the other hand, the number of layers evolves with time as layers exchange material and heat and potentially merge \citep{Vazan2015,Vazan2016}, with the two diffusive fluxes across the boundary being positively related to one another \citep{Moll2017}.  As such, even if planets formed with composition gradients that formed a large number of layers in a semi-convective staircase, the number of layers may have declined significantly as the planet subsequently evolved, reducing their impact on the present-day thermal structure of the planet.

For hot Jupiters, we lack direct evidence of the internal structure of the planets and must rely on comparisons with the solar system and the observed radii of the hot Jupiter population.  Whether metals are mixed into the envelope or condensed into a (homogeneous) core has a relatively modest effect on the radius of the planet \citep{Thorngren2016}.  On the other hand, many layers in a semiconvective staircase could inflate the radius by trapping heat in the planet's deep interior \citep{Leconte2012}.  This would be a poor solution to the radius inflation problem, as it would produce the same inflation in cooler giant planets, yet this is not observed \citep{Miller2011}.  Thus it is arguably most reasonable to assume the presence of a core sheathed in double-diffusive convective layers, but only a small number of layers in the outer envelope \citep[as seen by][for Jupiter and Saturn]{Vazan2016, Vazan2018}.  Such an arrangement would result in a warmer core than a two-layer model, but because heavy elements are much less thermally expansive than H/He, the effect on the planet's radius would be somewhat limited.

\subsection{Evolution Models} \label{sec:evolution}
For planets consisting entirely of adiabatic and/or isothermal layers with a continuous temperature profile, the thermal state of the planet is reduced to a single variable.  Typically this is the specific entropy of the envelope $s$.  A planet cools by releasing heat from the interior through its atmosphere, which depends on the thickness of the atmosphere (and therefore the mean molecular weight and gravity), its composition, and the incident flux from the star -- highly irradiated atmospheres release heat from the interior more slowly.  This can be specified as a total luminosity $L$ or an ``intrinsic temperature" $T_\mathrm{int}$, which are related via $L = 4 \pi R^2 \sigma_b T_\mathrm{int}^4$, where $\sigma_b$ is the Stefan-Boltzmann constant and R is the planet radius, \emph{usually} at the radiative-convective boundary.  Calculating the rate at which heat is released requires an atmosphere model that allows for both convective and radiative heat transport within the atmosphere down to the \hbindex{radiative-convective boundary} (RCB).  This procedure is outside the scope of this chapter, but readers are instead directed to \citep{Fortney2007} and the chapter of this book concerning radiative transfer.  The rate of change of total energy in the planet can be written as

\begin{equation}\label{eq:dedt}
    \frac{dE}{dt} = P_a + P_\mathrm{tide} + P_\mathrm{rad} - 4 \pi R^2 \sigma_b T_\mathrm{int}^4,
\end{equation}
where $P_a$ is the hot Jupiter anomalous heating power (see Sec. \ref{sec:radiusTrends} for estimates), $P_\mathrm{rad}$ is the heat generated by \hbindex{radioactive decays} in the planet, $P_\mathrm{tide}$ is the release of tidal energy within the planet.  This tidal heating could occur due to the damping of orbital eccentricity or obliquity between the planets rotational and orbital axes \citep[e.g.,]{Mardling2007, Jackson2008}, decaying to zero as the orbit circularizes and the rotation aligns to the orbit.

Heating from radioactive decays $P_\mathrm{rad}$ is commonly omitted from giant exoplanet studies, but work has been done to estimate its magnitude.  \citet{Nettelmann2011} and \citet{Lopez2012} discuss its inclusion in their sub-Neptune models, considering four major radioactive isotopes: $^{235}$U, $^{238}$U, $^{232}$Th, and $^{40}$K, whose half-lives are 0.7, 4.46, 14, and 1.25 Gyr, respectively.  The power generated over time is the product of the decay rate, decay energy, and the abundance of the isotope within the planet.  The first two are known very precisely, but the latter is difficult to estimate.  One option is to scale the isotopes with the modeled rock content in the same proportion as the protosolar abundances \citep[from e.g.,][]{Anders1989}.  Alternatively, \citet{Wang2020} suggest using the spectroscopic europium abundance of the host star as a proxy for $^{238}$U, $^{235}$U, and $^{232}$Th, since they are all refractory elements produced during r-process nucleosynthesis.  Carrying the former analysis through, however, \citet{Thorngren2023} found that the contribution of a hot Jupiter's luminosity from radioactive decays is extremely small and can be safely neglected.

Once a static structure model and the rate of change of energy in the interior has been calculated, calculating the change in envelope entropy with energy $ds/dE$ is a relatively simpler matter, requiring that one integrate the definition of temperature throughout the convective regions of the planet.  If the planet contains a core, then for that region a specific heat $c$ should be used and the temperature change determined by the change in temperature of the envelope at the core boundary with s.

\begin{equation} \label{eq:dsde}
    \left. 1 \middle/ \frac{ds}{dE}\right. = c M_c \frac{d T_c}{ds} + \int_{M_c}^M T(P,s) dm
\end{equation}

\begin{figure}[t]
    \centering
    \includegraphics[width=\textwidth]{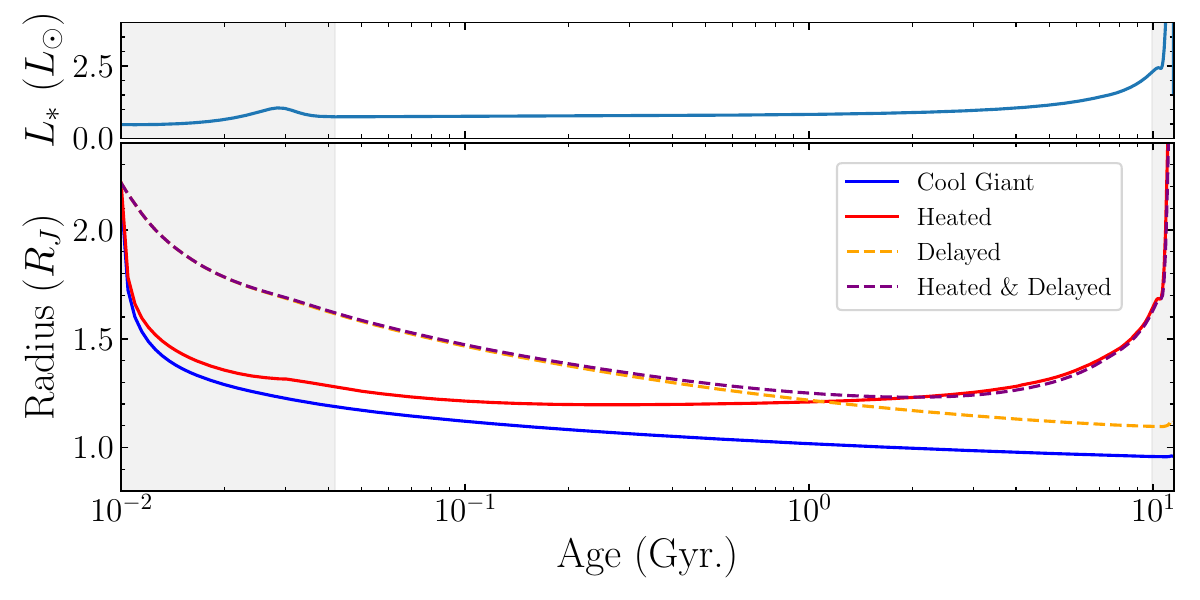}
    \caption{The radius evolution of a Jupiter mass planet under various assumptions about the thermal evolution, with the evolution of the parent (sun-like) star shown in the top panel, as this sets the incident flux on the planet.  Grey shading marks the pre- and post-main sequence. The solid blue line depicts a cool giant planet at 1 AU, which experiences no heating delayed cooling.  The solid red line shows a hot Jupiter heated entirely by deep heating, which consequently exhibits reinflation; the dashed orange line shows a hot Jupiter experiencing delayed cooling (here a constant factor of 20) and so does not reinflate but is larger at young ages.  Finally, the purple dashed line shows a combination of deep heating and reinflation, with the resulting slow initial cooling as well as reinflation.}
    \label{fig:evolutionExamples}
\end{figure}

Here, $c$ is the specific heat of the core, $M_c$ is the core mass, $T_c$ is the core temperature (as in this case it is assumed to be isothermal), $M$ is the total mass of the planet, $m$ indicates mass contained within a shell at the given point in the planet, and $T(P,s)$ is the temperature as a function of pressure $P$ and specific entropy $s$.  With Eqs. \ref{eq:dedt} and \ref{eq:dsde}, one can calculate the rate of change of the specific entropy as the product $ds/dE \times dE/dt$.  This may be integrated forward using a standard ODE (ordinary differential equation) integrator to obtain the \hbindex{thermal evolution} of the planet.  It's worth noting that for hot Jupiters these equations can be somewhat ``stiff'' (an ODE term outside the scope of this review), so it can be helpful to select an integrator that tolerates stiff ODEs such as backward differentiation formula (BDF) integrators.  Fig. \ref{fig:evolutionExamples} shows the radius evolution of a Jupiter-mass planet following this method for various assumptions about the anomalous heating and intrinsic temperature.

\section{Heating Depth and Reinflation} \label{sec:heatingDepthReinflation}
An important factor in how gaseous planets respond to heating is the depth at which the heat is deposited.  This is because convection can only move heat upwards -- hot fluid is less dense and rises through a colder medium, assuming a uniform composition.  As such, heat injected near the planet's surface can only slow cooling, not reverse it \citep[e.g.,][for the Ohmic dissipation case]{Wu2013}.  \citet{Komacek2017a} go further, showing that heat must be injected at least at the \hbindex{radiative-convective boundary} (RCB) in order to even slow the cooling; heat deposited above this point will be re-radiated out too quickly to affect the planet's radius or thermal evolution significantly.  While originally the RCB was thought to be very deep in the planet \citep{Guillot1996}, accounting for the planets' hot interiors indicates that the RCB will be at around 10 bar for most hot Jupiters \citep{Thorngren2019b}, moving to lower pressures with increasing metallicity and/or decreasing surface gravity.

A prediction of any model based on heating near the surface of the planet (see Sec. \ref{sec:mechanisms}) is therefore that planets will cool very slowly, but cannot grow larger (\hbindex{reinflate}) if the flux is later increased \cite[e.g.,][]{Wu2013,Komacek2020}.  \citet{Lopez2016} observe that this is testable by examining whether giant planets around evolved stars are disproportionately larger as a result of the increase in stellar luminosity.  \citet{Grunblatt2016,Grunblatt2017} find two such giant planets and observes that they are slightly larger than typical hot Jupiters would be without reinflation, though it would be helpful to find additional cases to verify the trend and potentially measure the rate at which reinflation occurs.

Post-main-sequence stars are not the only place to observe reinflation, however.  \citet{Hartman2016} observed a relationship between hot Jupiter radii and the fractional age of their parents stars -- the age relative to the main-sequence lifetime.  This could result, he argued, from the gradual brightening of the stars on the main sequence.  \citet{Thorngren2021} concur, observing that the degree of main-sequence brightening could indeed produce detectable increases in radius, and that these were consistent with the observed population.  However, they were not able to estimate the \emph{rate} that reinflation occurred at (which would be related to the deep-heating power), as main-sequence stellar brightening was too gradual.  For this, the rapid brightening of post-main-sequence stars remains the best place to look.

\section{Proposed Inflation Mechanisms}\label{sec:mechanisms}
One of the earliest proposed explanations for the inflated radii of hot Jupiters was that the model atmosphere \hbindex{opacity} was significantly lower than in real hot Jupiters \citep{Burrows2007,Burrows2008a}.  Such an error would cause models to cool too quickly, resulting in model planets that were too small.  The advantages of this explanation include its lack of need for any new physics or a source of deep heating.  However, it has a difficult time explaining the extremely large radii (up to $2 R_J$) that were subsequently discovered.  As such, although this may be an important source of theoretical uncertainty in hot Jupiter atmosphere models, it does not appear to be the primary driver of their anomalous radii.

Another solution suggested early on was that of \hbindex{tidal heating} \citep{Bodenheimer2001, Bodenheimer2003}.  Here, tides raised on a planet by the parent star would damp out the planet's eccentricity and spin-orbit obliquity, depositing the energy reduction into the planet's interior as heat.  However, this is a finite reservoir of energy and the tidal timescales for hot Jupiters are in general quite short \citep{Jackson2008}, so these processes would only be important in the earliest stages of a hot Jupiter's life.  Additionally, tidal forces scale strongly with semimajor axis and radius, not the incident flux as has been observed.  As such, while this may be important for some planets \citep{Millholland2020a} (not just hot Jupiters), it is not favored as a general solution \citep{Leconte2010, Thorngren2018}.  On the other hand, \citet{Arras2009} suggest that tidal heating may not arise from eccentricity or obliquity, but rather the tidal pull on the planet's thermal bulge \citep[][]{Socrates2013} which is offset from the sub-stellar point by winds.  In this \hbindex{thermal tides} model, energy is gradually pulled from the (enormous) reservoir of the planet's orbit and fed into the deep interior.  A key benefit of this model is that the deep heating could potentially reinflate the planet.  However, follow-up studies differ as to whether the scaling is able to match the correlation between radius and flux \citep{Sarkis2020, Thorngren2018}.

Several proposed explanations for radius inflation can be broadly categorized together as fluid dynamic in nature.  These share the common theme of arguing that the existing 1D atmospheric models are missing important physics that change the temperature-pressure profile of the deep atmosphere and suppress the release of heat from the interior.  One such model is the \hbindex{mechanical greenhouse} \citet{Youdin2010} proposal, in which the high-speed winds of the upper atmosphere generate turbulence beneath, dragging hot gas downward against the buoyancy force, producing a hotter and deeper RCB.  Such an effect would slow the cooling process of the hot Jupiter substantially, though it would not by itself be capable of reinflating planets that had already cooled.

More recently, \citet{Tremblin2017} proposed that the inflation effect may be resolved in higher-dimensional atmosphere models.  These 3D global circulation models (GCMs) are computationally expensive, especially when modeling the deep atmosphere where thermal timescales can be very long.  Instead, they solve for the steady-state solutions in 2D at various latitudes.  They find that \hbindex{temperature advection} longitudinally around the planet (steady-state flows, rather than turbulence) has the effect of burying heat deeper in the atmosphere than is indicated by 1D models.  The result is suppressed cooling, hotter interiors, and therefore substantially larger radii.  In fact, their model predictions were somewhat larger than the observed radii -- they suggest among other things that this may be the result of the 2D model that will not occur in a full 3D model.  \citet{Sainsbury-Martinez2019} studied this effect in 3D and supported their findings, suggesting that future studies use \emph{substantially} longer integration times and initialize their simulations with temperature-pressure profiles above equilibrium for faster convergence.

A somewhat related proposal to the fluid dynamical models relies on a common feature of GCM results for hot Jupiters: a fast eastward (super-rotating) jet stream \citep[e.g.,][]{Showman2011} carrying gas around the tidally locked planet.  In the presence of a planetary magnetic field, this jet may cause ions (especially Na) and free electrons to drift such that a current flows through the atmosphere from the polar regions to the equator \citep{Batygin2010}.  This current flows back to the poles through the deeper parts of the atmosphere, depositing energy as a result of the material's electrical resistivity.  This process, called \hbindex{Ohmic dissipation}, has been proposed as a source of the heating required to inflate hot Jupiters \citep{Batygin2010,Batygin2011}.

Originally it was thought that this would result in ultra-hot Jupiters inflating to the point that they overflow their Roche lobes, but subsequent studies disagree \citep{Spiegel2013}; \citet{Menou2012,Rauscher2013} found that magnetic braking can significantly reduce the velocity of the atmospheric winds, in turn reducing the effectiveness of Ohmic dissipation.  The peak efficiency of turning incident light energy into interior heating was found to occur at $1600$ K by \citet{Menou2012} and $1500-1600$ K by \citet{Rogers2014}.  Additionally, it is worth noting that this heating mechanism is fairly shallow, perhaps around 10 bars \citep{Rogers2014}, so it is not capable of causing reinflation \citep[][see Sec. \ref{sec:heatingDepthReinflation}]{Wu2013}.  \citet{Pu2017} investigated the process for hot Neptunes and found that Ohmic dissipation may operate in a similar manner in those objects as in hot Jupiters.  On the other hand, \citet{Huang2012} modeled the heating from Ohmic dissipation and instead argue that the total heating is \emph{not} sufficient to explain the radii observed among hot Jupiters.

\section{Observable Effects}
This section reviews some of the observation-based constraints that have been made to the nature of the hot Jupiter radius anomaly.  Because of the extremely large number of diverse observations made of hot Jupiters, it leaves to other chapters observations which do not directly relate to the inflationary mechanism, especially transmission and emission spectra (discussed elsewhere in this book).

\subsection{Population Trends in Radius}\label{sec:radiusTrends}
Given the large and growing population of observed hot Jupiters, a helpful approach to understanding them has been to consider their properties as a population, averaging out individual variations in metallicity or observational errors.  \citet{Laughlin2011} did this by calculating the radius anomaly $\mathcal{R}$, the radius minus the expected radius for an evolved metal-free planet, for the 90 transiting hot Jupiters known at the time.  They found that the anomaly was strongly related to the effective temperature, with $\mathcal{R} \propto T_\mathrm{eff}^{1.6\pm0.6}$, and argue that this appears consistent with the Ohmic dissipation model.  While they support the observation of a correlation between radius anomaly and incident flux, a key result is that the heating is not a fixed fraction of the incident flux, but one that itself increases with flux.

Following a similar thread, \citet{Thorngren2018} analyzed the then 318 hot Jupiters with masses and radii, directly including the internal heating power as a structure model parameter.  The degeneracy between heating and composition was resolved by imposing the \citet{Thorngren2016} mass-metallicity relation as a prior.  For individual planets, this doesn't resolve the degeneracy fully, but applied to a large population it becomes possible to measure heating trends.  They found that the \hbindex{heating efficiency} as a fraction of flux $\epsilon$ rose to a peak of about 2.5\% at 1570 K ($F=1.38\times10^9$ \ergscm) and then declines to below 1\% for the ultra-hot ($>2000$ K) planets.  They found a good fit using a Gaussian in log-flux to predict the heating:
\begin{equation}\label{eq:epsilonThorngren}
    \epsilon(F) = \epsilon_\mathrm{max} \exp \left(- \frac{(\log(F) - F_0)^2}{2w} \right)
\end{equation}

Here, the amplitude is $\epsilon_\mathrm{max} = 2.37_{-.26}^{1.3}$, the center is at $F_0 = .14_{-.69}^{+.60}$, and the width is $w = .37^{+.038}_{-.059}$.  Interestingly, there are hints of this decline in heating efficiency for the ultra-hot Jupiters in \citet{Laughlin2011} (that paper's Fig. 2.), but there were too few such planets known at the time for this to have been statistically significant.

\citet{Sarkis2020} conducted a similar study, seeking to extract the heating efficiency from the population of hot Jupiters with a Bayesian hierarchical model.  Unlike \citet{Thorngren2018}, they included a number of rigorous checks including whether they could retrieve the modeled value from a synthesized population (they did), checks for prior sensitivity (their results were insensitive), and more extensive comparisons with predicted heating from various models.  They too found a peak in heating efficiency of about 2.5\%, but at a higher temperature of $\sim1860$ K; this difference does not appear to be due to choice of prior or the statistical framework, but may be a difference in atmosphere models.  They present several parameterizations for estimating $\epsilon$; here is the Gaussian model (in terms of $T_{eq}$) for reference and comparison:
\begin{equation}
    \epsilon(T_\mathrm{eq}) = \epsilon_\mathrm{max} \exp \left( -
        \frac{\left(T_\mathrm{eq} - T_\mathrm{eq,0}\right)^2}{2s^2}
    \right)
\end{equation}
where $T_\mathrm{eq}$ is the equilibrium temperature, $\epsilon_\mathrm{max}$ is the peak heating efficiency, $T_\mathrm{eq0}$ is the temperature where the peak is located, and $s$ is the width of the Gaussian shape.  Using a linear prior on $\epsilon$ (they also tried using a log prior), they found $\epsilon_\mathrm{max}=2.46_{-.24}^{+.29}$, $T_\mathrm{eq0} = 1862_{-61}^{+67}$, and $s=508_{-48}^{+66}$.  Comparing their inferred heating with model predictions, they agreed that the Ohmic heating found by \citet{Huang2012} was not sufficient to explain the radii, but that thermal tides \citep{Socrates2013} and temperature advection \citet{Tremblin2017} potentially could.  That said, they did still find that the latter two, especially thermal tides, over-predicted the heating for ultra-hot Jupiters.

For the assumptions made by both \citet{Thorngren2018} and \citet{Sarkis2020} (deep heating), the planet rapidly approaches an equilibrium state in which the intrinsic luminosity of the planet is equal to the anomalous heating.  This  implies a direct relationship with the intrinsic temperature of the planet:
\begin{equation}\label{eq:tint}
    T_\mathrm{int} = \left(\frac{\epsilon F}{4 \sigma}\right)^\frac{1}{4} = \epsilon^{1/4} T_\mathrm{eq}
\end{equation}
where $\sigma$ is the Stefan-Boltzmann constant.  Remember that $\epsilon$ depends on flux (or $T_\mathrm{eq}$), so this is not to be understood as a linear relationship between $T_\mathrm{int}$ and $T_\mathrm{eff}$.  Still, using one of the formulas for $\epsilon$ allows one to calculate the $T_\mathrm{int}$ relatively easily \citep[e.g.,][]{Thorngren2019}, which can be particularly helpful for studies where an estimate of it is needed; such applications are discussed in the next section.

\subsection{Intrinsic Temperature and Radiative-Convective Boundary} \label{sec:tintRCB}
A hot interior has the potential to substantially affect the structure and heat flow through the planet's atmosphere.  The \hbindex{intrinsic temperature} $T_{\mathrm int}$ (see also Sec. \ref{sec:evolution}) is a way to describe the energy entering the atmosphere from below.  It participates in the heat balance of the atmosphere as:
\begin{equation}
    E_\mathrm{out} = E_\mathrm{irrad} + E_\mathrm{interior}
\end{equation}
\begin{equation} \label{eq:teff}
     T_\mathrm{eff}^4 = T_\mathrm{eq}^4 + T_\mathrm{int}^4
\end{equation}

That is, the heat exiting the planet is the sum of the incoming light energy $E_\mathrm{irrad}$ plus the heat from the interior $E_\mathrm{interior}$.  Astronomers often rewrite this in terms of temperatures (Eq. \ref{eq:teff}) due to the popularity of the equilibrium temperature and effective temperatures.  Note that $T_\mathrm{int}$ is the \emph{intrinsic temperature} and not the interior temperature, which varies with pressure!  Nowhere in the interior will the actual temperature be as low as the intrinsic temperature.  For cool giant planets, $T_\mathrm{int}$ is initially large and drops over time as the planet cools; however, for hot Jupiters the anomalous heating may replace some or all of the heat going out, resulting in a steady-state intrinsic temperature in the hundreds of kelvin \citep{Thorngren2018,Sarkis2020}.  On the other hand, if the large radii are caused by delayed cooling the intrinsic temperature would be by definition \emph{smaller} than that predicted by existing atmosphere models.

The intrinsic temperature can have detectable impacts on the observable atmosphere \citep[see][]{Fortney2020, Gao2020}; for example \citet{Sing2019} argue that the presence of Fe and Mg in the atmosphere of WASP-121 (which they observed with HST) requires a $T_\mathrm{int}$ of at least 500 K to avoid the atoms condensing in the atmosphere too deep to be observed.  This is consistent with the deep heating case -- \citet{Sarkis2020} calculate a $T_\mathrm{int}=800$ K for that planet.  The presence of absence of these cold traps is therefore a useful diagnostic of the intrinsic temperature of some planets \citep{Parmentier2016,Fortney2020}.  The atmospheric chemistry may also be affected by $T_\mathrm{int}$, as higher values result in a higher \hbindex{eddy mixing parameter} $k_{zz}$, pushing the chemical abundances further from equilibrium.  The intrinsic temperature is also an important input in global circulation models (GCMs) of giant planets \citep[e.g.,]{May2021,Steinrueck2021,Lee2022}, though in that field it is referred to as internal temperature.

The intrinsic temperature and equilibrium temperature also directly affect the location of the \hbindex{radiative-convective boundary} (RCB, see Sec. \ref{sec:heatingDepthReinflation}): a higher $T_\mathrm{eq}$ pushes the RCB to higher pressures, whereas a high $T_\mathrm{int}$ pushes it to lower pressures \citep[e.g.,][]{Guillot1996,Fortney2007,Thorngren2019}.  When hot Jupiters were initially discovered, it was believed that their very high equilibrium temperatures would move the RCB very deep into the planet, at around ~1 kilobar \citep[e.g.,][]{Guillot1996}.  However, these models were based on RV discoveries, which had not yet revealed the radius anomaly problem.  If large radii indeed imply large intrinsic temperatures, \citet{Thorngren2019b} and \citet{Sarkis2020} both find that the RCB is pushed back up to 1-100 bars, depending on the planets' exact equilibrium temperature and surface gravity.  As \citet{Sarkis2020} note, heat must be deposited below the RCB to have a significant impact on the radius \citep{Komacek2017}, so a shallower RCB imposes less stringent requirements on the pressures of heat deposition.  They argue further that the same argument could be applied to the temperature advection mechanism of \citet{Tremblin2017}.

\subsection{Magnetic Fields} \label{sec:magneticfields}
\begin{figure}[t]
    \centering
    \includegraphics[width=\textwidth]{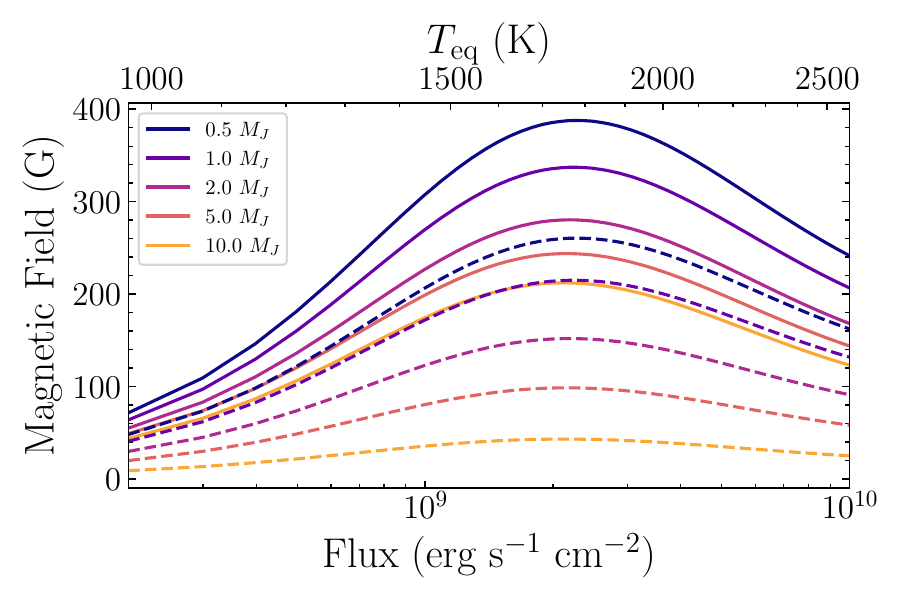}
    \caption{The magnetic field strength at the dynamo surface (solid) and polar surface (dashed) for a hot Jupiter as a function of incident flux for various masses (color) from the \citet{Christensen2009} scaling relations (Eq. \ref{eq:bfield}). Here the intrinsic temperatures from Eqs. \ref{eq:epsilonThorngren} and \ref{eq:tint} and radii from Eq. \ref{eq:massRadiusFluxThorngren} are used.  The equivalent zero-albedo equilibrium temperature for efficient redistribution of heat is shown as the top x-axis.}
    \label{fig:bfield}
\end{figure}

The \hbindex{magnetic field}s of exoplanets are poorly understood due to the difficulty in making direct measurements of them, but are thought to relate to the rotation rate, intrinsic temperature, or both \citep{Reiners2010}.  Hot Jupiters are expected to be tidally locked \citet{Jackson2008}, so even on their short periods the rotation rates will be lower than those of Jupiter and Saturn.  On the other hand, their intrinsic temperatures appear to be substantially higher (Sec. \ref{sec:tintRCB}).  Assuming that hot Jupiter radii are explained by injected heat, \citet{Yadav2017} applied the \citet{Christensen2009} scaling relations to infer hot Jupiter magnetic field strengths.  The mean field on the dynamo surface from \citet{Christensen2009} was:
\begin{equation}\label{eq:bfield}
    B_\mathrm{mean}^\mathrm{dyn} = 4800 \left(\frac{M}{M_\odot}\right)^{1/6}
    \left(\frac{L_\mathrm{int}}{L_\odot}\right)^{1/3} \left(\frac{R}{R_\odot}\right)^{-7/6} \mathrm{G}
\end{equation}
where $M$ is the planet mass, $L_\mathrm{int}$ is the planet's intrinsic luminosity (use the formulas from Sec. \ref{sec:tintRCB} and $L_\mathrm{int} = 4 \pi \sigma T_\mathrm{int}^4$), and R is the planet's radius.  For hot Jupiters, this results in polar surface field strengths of roughly 50-200 G (see Fig. \ref{fig:bfield}), depending strongly on the incident stellar flux and planet mass.  Such fields would be substantially stronger than those seen in Jupiter or Saturn, and could have impacts on the atmospheric flows \citep[e.g.,][]{Beltz2023}.  These results are a reasonable estimate for what hot Jupiter fields might be, but they could be improved on by considering how the rotation rates factor in \citep[e.g.,][for stars]{Jeffries2011} and a more careful treatment of where the dynamo surface is.

An important angle for observationally constraining hot Jupiter magnetic fields is the interactions between the fields of the planet and its parent star.  The \hbindex{Ca II K line} is known to be a tracer of magnetic activity in stellar chromospheres \citep[e.g.,][]{Schrijver1989}, and so it has been hypothesized as a potential avenue for measuring star-planet magnetic interactions \citep[see][]{Lanza2008,Poppenhaeger2019}, though the point of peak activity may not be directly facing the planet.  Subsequent observations have found cases where the Ca II K line varies at the orbital period of the planet \citep[e.g.,][]{Shkolnik2003, Shkolnik2005}, supporting the hypothesis that it is driven by star-planet interactions.  An important development was made by \citet{Cauley2019}, who calculated the magnetic field strengths implied by the Ca II K line modulation amplitudes for four hot Jupiters and found strong magnetic fields with surface strengths of 20-120 G, consistent (if possibly slightly lower) with the intrinsic temperature-based scaling laws used by \citet{Yadav2017}.  Thus it provides observational support for the hypothesis that hot Jupiter magnetic fields are significantly stronger than those of cooler giants like Jupiter and Saturn.

The magnetic field of the planet is also thought to have an effect on the mass loss processes observed on some hot Jupiters \cite[see][for an overview]{DosSantos2021}.  \citet{Adams2011} show how magnetic fields can inhibit the rate of XUV-driven mass loss from giant planets with even relatively modest magnetic fields ($B > 1$ G); \citet{Owen2014} agree and further argue that planetary magnetism will largely prevent mass loss from the night-side by inhibiting flows from the day-side (where the XUV irradiation is absorbed) around the planet.  Additionally, the resulting outflows will be sculpted by the planetary magnetic fields; most importantly, escaping plasma follows the magnetic fields lines until it reaches the planet's Alfvén surface, preventing mass from escaping from the regions away from the poles where the field lines never cross the Alfvén surface \citep{Adams2011, Carolan2021}.  Using these patterns, \citet{Schreyer2023} argues that the blue-shifting observed from escaping gas is tentative evidence for a magnetic field weak enough ($< 0.3$ G) that gas can escape from the night side, as day-side escape would be red-shifted.  The growing number of observations of gas escaping from hot Jupiters makes this a promising area for further theoretical work as well.

The use of radio observations has been suggested as a more direct approach for measuring the magnetic fields of hot Jupiters \citep{Zarka2007, Weber2018} based on the successful observation of the aurorae of Jupiter in radio \citep[see][]{Zarka1998}.  Such searches would be made easier if the planets had a moon as a source of ions to supply the aurorae \citep{Narang2023}, but such moons are more likely for young distant objects than for hot Jupiters, where they would not be stable on long timescales \citep{Barnes2002}.  Searches have been conducted mainly for young, widely-separated planets \citep[e.g.,][]{Bastian2018, Shiohira2024, Turner2024}.  These searches have not yet identified signals clearly attributable to the planet, but efforts are ongoing and still promising due to the successes in observing such signals from brown dwarfs \citep[e.g.,][]{Kao2018}.

\subsection{Tidal Effects}
Hot Jupiters on particularly close orbits are likely to be tidally distorted into a non-spherical shape by their parent stars.  Unlike cooler giant planets, whose rotation can warp them into oblate spheroids, hot Jupiters are pulled into mostly-prolate \citep[e.g.,][]{Li2010} tri-axial ellipsoids, where each of the three axes (sub-stellar, polar, and prograde) has a different radius.  Estimating the self-consistent equilibrium shape theoretically is numerically expensive, but has been done for some of the most tidally-affected hot Jupiters by \citet{Wahl2021} using the concentric MacLaurin spheroid method. For moderately distorted planets, a good and much faster approximation of the shape can instead be achieved using the Love number $k2$ \citep{Love1909,Leconte2011,Akinsanmi2019, Hellard2019}, which is obtained from a static structure model via a relatively simple ODE \citep[e.g.,][]{Buhler2016}:
\begin{equation}
    r \frac{d\eta_2(r)}{dr} = 6 + \eta_2(r) - \eta_2(r)^2  - 6 \frac{\rho(r)}{\rho_m(r)}(\eta_2(r)+1)
\end{equation}
\begin{equation}\label{eq:k2}
    k2 = \frac{3-\eta_2(R)}{2+\eta_2(R)}
\end{equation}
where $\eta_2$ is an intermediate quantity, $r$ is the radius in the planet, $R$ is the total radius of the planet, $\rho(r)$ is the density at radius $r$, and $\rho_m(r)$ is the average density within the sphere of radius $r$.  The ODE is to be integrated from the initial conditions radially outwards from $\eta_2(0)=0$, so that $\eta_2(R)$ may be plugged into Eq. \ref{eq:k2} to obtain $k2$.  Determining the Love number for hot Jupiters is of particular interest because, as a measure of how concentrated the planet's density is towards the center, it can break the degeneracy between composition and specific entropy without any assumptions about the evolutionary history \citep[e.g.,][]{Hellard2019}!

Observation of the tidal distortion is challenging but possible, and several avenues are available.  First, the shape affects the shape of the transit by extending the amount of time spent in ingress and egress and a subtle decrease in transit depth towards mid-transit as the long axis of the planet faces the viewer \citep{Leconte2011}.  The latter effect is difficult to separate from the effects of limb darkening, but the ingress and egress duration should be measurable in favorable cases \citep{Akinsanmi2019}.  \citet{Barros2022} have claimed a 3-sigma detection of the shape of WASP-103 b using this approach.  An alternative approach is a careful treatment of the the phase curve of the planet, which is affected by the shape even more strongly than the transit is \citep{Akinsanmi2023}; however, this signal must be carefully disentangled from others, including the planet temperature vs latitude, stellar deformation by the planet, and Doppler beaming of the emitted light.

A case of particular interest is that of HAT-P-13 b and c.  Using a secular dynamics analysis, \citep{Batygin2009} showed that the system fairly quickly reaches a steady-state in which gravitational interactions between b and c transfer orbital energy into the eccentricity of b, which is then damped by circularization tides onto a constant value of $e_b$.  More importantly, the two orbits precess together, and while the details of the secular dynamics arguments are too involved to go into here, the result is that the eccentricity of b is directly related to its value of $k_2$!  Follow-up by \citep{Buhler2016} used HST secondary transits to determine an eccentricity of $e=0.007 \pm 0.001$, implying ${k}_{2}={0.31}_{-0.05}+^{+0.08}$, which corresponds to a most likely core mass of $11 M_\oplus$ and a 1-sigma upper limit of $25 M_\oplus$, reasonable numbers for the $.85 M_J$ object.  This is so far the only constraint on an exoplanet's $k_2$ precise enough to be useful for constraining its interior structure, but it seems likely that more will be measured, especially as more long-period companions of known exoplanets are discovered.

\section{Conclusion}
The hot Jupiter radius inflation problem continues to  be one of the longest-standing open questions in exoplanet science; however, significant progress has been made over the last two decades in collecting observational evidence for its cause.  To review the key points of evidence from previous sections:
\begin{enumerate}
    \item The entire population of hot Jupiters shows signs of inflation; the lower-radius examples are consistent with higher metallicity, not a lack of inflation \citep{Laughlin2011}.
    \item The degree of radius inflation is correlated with incident stellar flux, not period or semimajor axis \citep{Weiss2013}.
    \item The fraction of flux used to inflate the planet initially increases with flux, then decreases for the ultra-hot Jupiters; it does not appear to depend strongly on mass \citep{Thorngren2018,Sarkis2020}.
    \item Hot Jupiters show signs of reinflation around both main-sequence stars and post-main-sequence stars \citep{Grunblatt2016, Grunblatt2017, Hartman2016, Thorngren2021}.
    \item The absence of cold traps on WASP-121 b indicates a high intrinsic temperature \citep{Sing2019}.
    \item Star-planet magnetic field interactions tentatively suggest strong planetary magnetic fields \citep{Cauley2019}, in turn implying a high intrinsic temperature \citep{Yadav2017}.
    \item Young hot Jupiters appear to be slightly larger than their older analogues \citep{Thorngren2021}.
\end{enumerate}

Comparing these observations with the proposed inflation mechanisms in Sec. \ref{sec:mechanisms}, it is evident that there is a tension between them.  The presence of a uniform inflation effect strongly related to incident flux suggests that the inflation mechanism is closely tied to atmospheric dynamics, such as Ohmic dissipation or temperature advection.  However, these mechanisms are not able to explain the reinflation of hot Jupiters seen in several studies.  Though thermal tides models are able to bring about reinflation, the heating predicted by them does not appear to match the radii of ultra-hot Jupiters \citep{Sarkis2020}, seeming to overinflate them \citep{Thorngren2018}.  One possibility is that we are missing an important effect that limits the impact of thermal tides at very high equilibrium temperatures -- perhaps magnetic breaking prevents these objects' hot spots from moving as far.

Alternatively, it may be necessary to consider multiple mechanisms operating at the same time.  If both thermal tides \emph{and} Ohmic dissipation contribute significantly, for example, they might match the observations by having Ohmic dissipation regulate heat flow out of the planet while modest thermal tides allow the planet to reinflate.  The key point here is that it isn't necessarily the case that any one inflation mechanism solve the entire problem on its own -- they may be acting in concert with one another.

To conclude, here are some promising areas of investigation.  On the observational side, the discovery of more hot Jupiters is always helpful for doing population statistics, but those at very young ages or around post-main sequence stars will be particularly valuable as constraints on the radius evolution of hot Jupiters.  For individual planet studies, any atmosphere observations that can measure or constrain the intrinsic temperature \citep[as in][]{Sing2019} are of particular interest, as would constraints on the atmospheric metallicity, as these help to narrow down the range of plausible interiors.  Observations of the magnetic field strengths of hot Jupiters is similarly helpful, be they through star-planet interactions, direct radio observation, or some other method.  Finally, a sufficiently precise observation of the prolate shape of a hot Jupiter (or equivalently its Love number) would provide a valuable new angle of constraint on the deep interior.

For theoretical work, a particularly noteworthy need is a revisitation of the thermal tides model, which has not been substantially developed since \citet{Socrates2013}.  Our improved understanding of hot Jupiter circulation, including the day-night contrasts and hot-spot offsets, warrant further work.  Likewise, GCMs extending down into the deep atmosphere on long enough integration times for various masses and instellation would be computationally expensive but extremely valuable in addressing the temperature advection mode.  Considering also the effect of magnetic fields worsens the computational expense, but may be necessary to understand how these revised atmospheric flows interact with Ohmic dissipation.  Finally, the rapidly enlarging sample of observed hot Jupiters and recent developments in H/He equations of state make it worth recalculating the trends in heating with incident flux with particular focus on the cooling and reinflation rates.

Between the abundance of new observations and the ongoing theoretical work to understand them, this is a very exciting time for the study of hot Jupiters.  The coming years will likely see further dramatic advancements in understanding of these planets.

\section{Cross-References}
\begin{itemize}
    \item Radiative Transfer for Exoplanet Atmospheres - Heng \& Marley
    \item Exoplanet Atmosphere Measurements from Transmission Spectroscopy and Other Planet Star Combined Light Observations - Kreidberg
    \item Tidal Star-Planet Interactions: A Stellar and Planetary Perspective - Mathis
\end{itemize}

\begin{acknowledgement}Thanks go to the Allan C. and Dorothy H. Davis Postdoctoral Fellowship at Johns Hopkins University for support, and to Stephen Schmidt for extensive proofreading and readability suggestions.
\end{acknowledgement}

\bibliographystyle{spbasicHBexo}
\bibliography{main}
\end{document}